\documentclass[12pt]{article}
\usepackage{color}
\usepackage{epsfig}
\usepackage{amsmath}
\usepackage{hhline}
\usepackage{amsfonts,amssymb}
\usepackage{times}

\graphicspath{{./eps/}}
\DeclareGraphicsExtensions{.eps,.eps.gz,.ps,.ps.gz}
\newlength{\dinwidth}
\newlength{\dinmargin}
\newcommand{\hdick}{\noalign{\hrule height1.4pt}}

\newcommand{\GeV} {\mathrm{GeV}}

\newcommand{\fb}  {\mathrm{fb}}
\newcommand{\fbi} {\mathrm{fb}^{-1}}

\newcommand{\cL } {{\cal L}}
\newcommand{\cP } {{\cal P}}
\newcommand{\cB } {{\cal B}}

\def\pythia  {{\sc Pythia}}

\def\tesla  {{\sc Tesla}}

\def\ee{e^+e^-}
\def\ti    {\tilde}

\def\stau  {{\ti\tau}}

\def\sell  {{\ti\ell}}

\def\sl    {{\ti\ell}}
\def\cx    {\ti {\chi}}

\def\cp    {\ti {\chi}^+}

\def\nt    {\ti {\chi}^0}

\def\smu   {{\ti\mu}}

\def\smur  {{\ti\mu}_R}

\def\smurm {{\ti\mu}^-_R}

\def\smurp {{\ti\mu}^+_R}

\def\se    {{\ti e}}
\def\sel   {{\ti e}_L}
\def\ser   {{\ti e}_R}

\def\serm  {{\ti e}^-_R}

\def\serp  {{\ti e}^+_R}

\def\snl  {{\ti\nu}_l}

\def\tstau {\theta_{\ti\tau}}

\newcommand{\mstau}[1] {m_{\ti \tau_{#1} }}
\newcommand{\mnt}[1]   {m_{\ti \chi^0_{#1} }}

\def \Eslash {E \kern-.75em\slash }
\def \Mslash {M \kern-.5em\slash }

\newcommand{\beq}{\begin{equation}}
\newcommand{\eeq}{\end{equation}}
\newcommand{\bea}{\begin{eqnarray}}
\newcommand{\eea}{\end{eqnarray}}

\newcommand{\fig}[1]{fig.~\ref{#1}}

\begin{document}

\title{ 
  \begin{flushright} \normalsize
    { LC--PHSM--2003--071} \\
    August 2003 \\[1em]
  \end{flushright} 
  \LARGE\bfseries
  Study of Sleptons at a Linear Collider -- Supersymmetry  
  Scenario SPS~1a\footnote{
    Contribution to the Extended ECFA/DESY Study on Physics and
    Detectors for a Linear Electron-Positron Collider (2001 -- 2003)}
    }
\author{\Large Hans-Ulrich Martyn\footnote{email: martyn@mail.desy.de}
  \\[.5ex]  
  { \itshape 
     I. Physikalisches Institut, RWTH Aachen, Germany}}   
\date { }
 
\maketitle
\thispagestyle{empty}

\begin{quote}   \small
The properties of the light charged sleptons in
$\ser\ser$, $\smur\smur$ and $\stau_1\stau_1$ production  
are studied within the mSUGRA scenario SPS~1a.
Various method are presented to determine very precisely masses,
angular distributions, spin and couplings as well as the $\stau_1$
polarisation and mixing parameter in the $\stau$ sector.
The analysis is based on a complete simulation of signal and
background reactions assuming experimental conditions of the 
\tesla \ $\ee$ Linear Collider.

\end{quote}

\section{Introduction}

This note describes, for a particular bench mark model, how masses and
couplings of scalar leptons can be accurately measured at a future
$\ee$ linear collider \tesla~\cite{tdr}. 
Beam conditions are chosen such as to stay below production thresholds 
of the heavier sleptons, the assumed luminosity corresponds to about one year
of running at design parameters. 
Details of the experimental simulation, analysis
and background information on kinematics and theory
are given in sketchy form.

\paragraph{Particle spectrum} 
 mSUGRA SPS~1a scenario~\cite{spsmodels}  
 (parameters 
 $m_0 =  100\;\GeV$, $m_{1/2} = 250\;\GeV$, $A_0 = -100\;\GeV$, 
 $\tan\beta = 10$, sign$\,\mu\, + $)
 shown in \fig{mass_spectrum}. 
 Cascade decays of $\nt_2$, $\cx_1^\pm$ lead to abundant $\tau$ final
 states.

\paragraph{Event generation} \pythia~6.2~\cite{pythia} including 
 beam polarisations $\cP_{e^\pm}$, 
 QED radiation, beamstrahlung \`a la {\sc Circe}~\cite{circe}, 
 %beam polarisations $\cP_{e^\pm}$, 
 $\tau$ decays and $\tau$ polarisation treated by {\sc Tauola}~\cite{tauola}

 \paragraph{Operating conditions}
 $\sqrt{s} = 400~\GeV$, beam polarisations  
 $\cP_{e^-} = +0.80$,  $\cP_{e^+} = -0.60$,
 integrated luminosity  $\cL = 200~\fbi$ \\ $\Rightarrow$
 efficient suppression of SUSY ($ \snl,\ \nt_2,\ \cx^\pm_1$) 
 and SM background

 \paragraph{Detector \`a la TDR \cite{tdr}} Simulation based on
 {\sc Simdet}~4.02~\cite{simdet}, acceptance $\theta > 125$~mrad, \\
 $e, \, \gamma$ veto $\theta > 4.6$~mrad

\paragraph{Slepton production} 
\begin{eqnarray}
  e^+_L e^-_R & \to & \smur\smur \qquad (120\,\fb)
  \label{smurproduction}
  \\
  e^+_L e^-_R & \to & \ser\ser \qquad \ (550\,\fb)  
  \label{serproduction}
  \\
  e^+_L e^-_R & \to & \stau_1\stau_1 \qquad \ \ \ (140\,\fb) 
  \label{stauproduction}
\end{eqnarray}

\paragraph{Background production} 
      SUSY bkg \quad 
      $e^+_L e^-_R \to \ser\sel\, (75\,\fb), \ 
                       \nt_2\nt_1\, (20\,\fb), \ 
                       \cp_1\cp_1\, (5\,\fb)$  \\      
      SM bkg  \quad
      $e^+_L e^-_R \to W^+ W^-\, (1000\,\fb)$  
      [$B_{W\to\ell\nu} = 0.107$],
      \quad 
      $\ee\to \ell^+\ell^- (\gamma)$ %,  $e^+e^-\ell^+\ell^-$ 
      negligible
      \\  \phantom{xxxxxxxxi}
      $\gamma\gamma$ bkg negligible \quad
      $\sigma(\ee\to e^+e^-\tau^+\tau^-) = 4.5\cdot 10^5~\fb$,
      acceptance $< {\rm few} 10^{-6}$ 

\paragraph{Slepton decays} 
\begin{eqnarray}
  \sell^- & \to & \ell^- \, \nt_1      \label{sldecay}
\end{eqnarray}
Flat lepton energy spectrum with endpoints $ E_{+/-}$ related to
slepton and neutralino masses
\begin{eqnarray}  
  E_{+/-} & = &
        %\frac{m_{\sl}}{2}  
        %\left ( 1 - \frac{m_{\cx}^2}{m_{\sl}^2} \right ) 
        %\gamma \, (1 \pm \beta)   \\[.5ex]
        %\frac{1 \pm \beta}{\sqrt{1-\beta^2}}   \\[.5ex]
        \frac{\sqrt{s}}{4} 
        \left ( 1 - \frac{m_{\cx}^2}{m_{\sl}^2} \right ) 
        \, (1 \pm \beta)   \\[.5ex]
        m_{\tilde{l}} & = &
        %\frac{\sqrt{s}}{E_{-}+E_{+}}\,\sqrt{E_{-}\cdot E_{+}} \\[.5ex]
        \frac{\sqrt{s}}{E_{-}+E_{+}}\,\sqrt{E_{-}\,  E_{+}} \\[.5ex]
        m_{\cx} & = & m_{\tilde{l}} \,
          \sqrt{1 - \frac{E_{-}+E_{+}}{\sqrt{s}/2}} 
\end{eqnarray}

\paragraph{{\boldmath $\ell^+\ell^-$} momentum correlations \cite{feng}}

  \setlength{\unitlength}{1mm}
  \begin{picture}(150,0)(0,0) 
    \thicklines
    \put(80,-15){
      \put(0,0){\line(1,0){18}}
      \put(0,0){\line(-1,0){18}}
      \put(5,1){{\small $\theta$}}
      \put(20,-1){{\small $e^-$}}
      \put(-23,-1){{\small $e^+$}}
      {\color{red}
        \put(1.5,0){\vector(1,1){15}}
        \put(1.5,0){\vector(-1,-1){15}} 
        \put(11,7){{\small $\smu^+$}}
        \put(-7,-11){{\small $\smu^-$}}
        \put(11,7){{\small ${\smur}^+$}}
        \put(-7,-11){{\small ${\smur}^-$}}
        }
      {\color{blue}
        \put(0,0){\vector(1,2){4.5}}
        \put(0,0){\vector(-2,-1){14.5}}  
        \put(-2,5){{\small $\mu^+$}} 
        \put(-13,-4){{\small $\mu^-$}}
        }
      {\color{green}
        \put(14.5,14.5){\line(-2,-1){11}}
        \put(-15.3,-15.3){\line(0,1){8}}
        \put(6,13){{$\cx$}}
        \put(-19,-11){{$\cx$}}
        }
      }
  \end{picture}
\noindent
$m_{\nt}$ known: \quad energy-momentum conservation constrains \\
construction of
kinematically allowed minimum mass $m_{\rm min}(\sell)$ \\
sharp peak around true mass $m_{\sell}$ 

\noindent
$m_{\nt}$ and $m_{\sell}$ known: \quad reconstruct $\sell$ direction
$\theta$ \\
two-fold ambiguity, 
false solution flat in $\cos\theta$

\section{Properties of smuons}

\paragraph{ Selection criteria} 
  $e^+_L e^-_R \to \smur\smur$ at $\sqrt{s}=400~\GeV$
  \footnotesize \scriptsize
  \begin{verbatim} 
 nr  cut                              min       max      eff 
 ------------------------------------------------------------
  0. n_ch, n_gamma in accept.           2         0     0.815
     n_charge in acceptance             2         2     0.989
     n_gamma  in acceptance           5.0         0     0.851
     n_veto   in acceptance           5.0         0     0.960
  1. Q*cos_theta_l                 -0.900     0.750     0.693
  2. delta phi_(l+,l-) (deg)                  160.0     0.555
  4. p_T event (GeV)                 10.0               0.553
  5. p_miss cos_theta_(l+l-)                  0.900     0.537
  7. lepton energy (GeV)              2.5     150.0     0.537
 10. m_recoil + 0.4*m_ll (GeV)      240.0               0.537

  efficiency     0.5365    

  lepton energy  E_min 16.52       E_max 93.30    
  \end{verbatim}
  \normalsize

\begin{figure}[htb] \centering
  \mbox{\hspace{-5mm}
  \epsfig{file=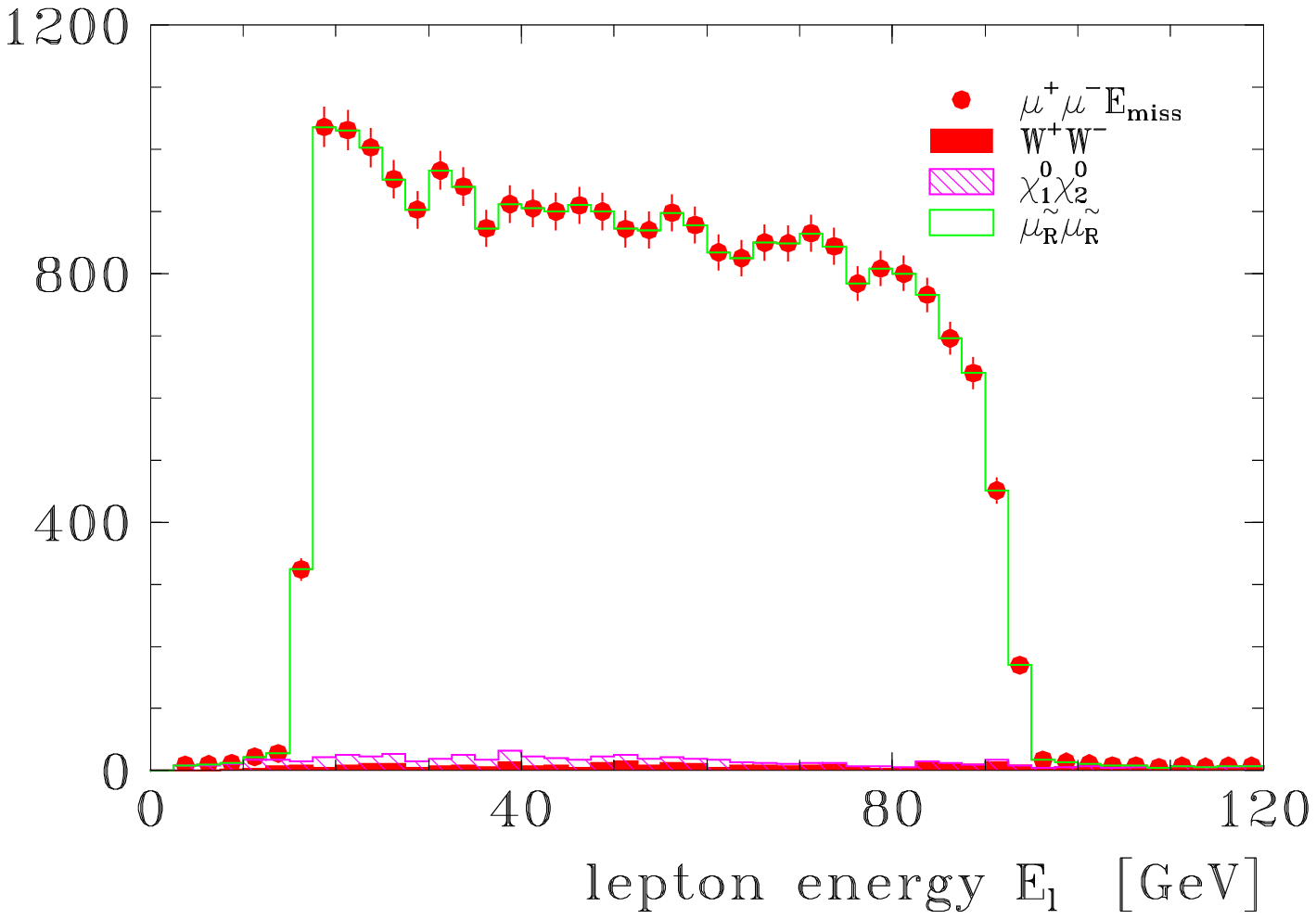,width=.52\textwidth} 
  \epsfig{file=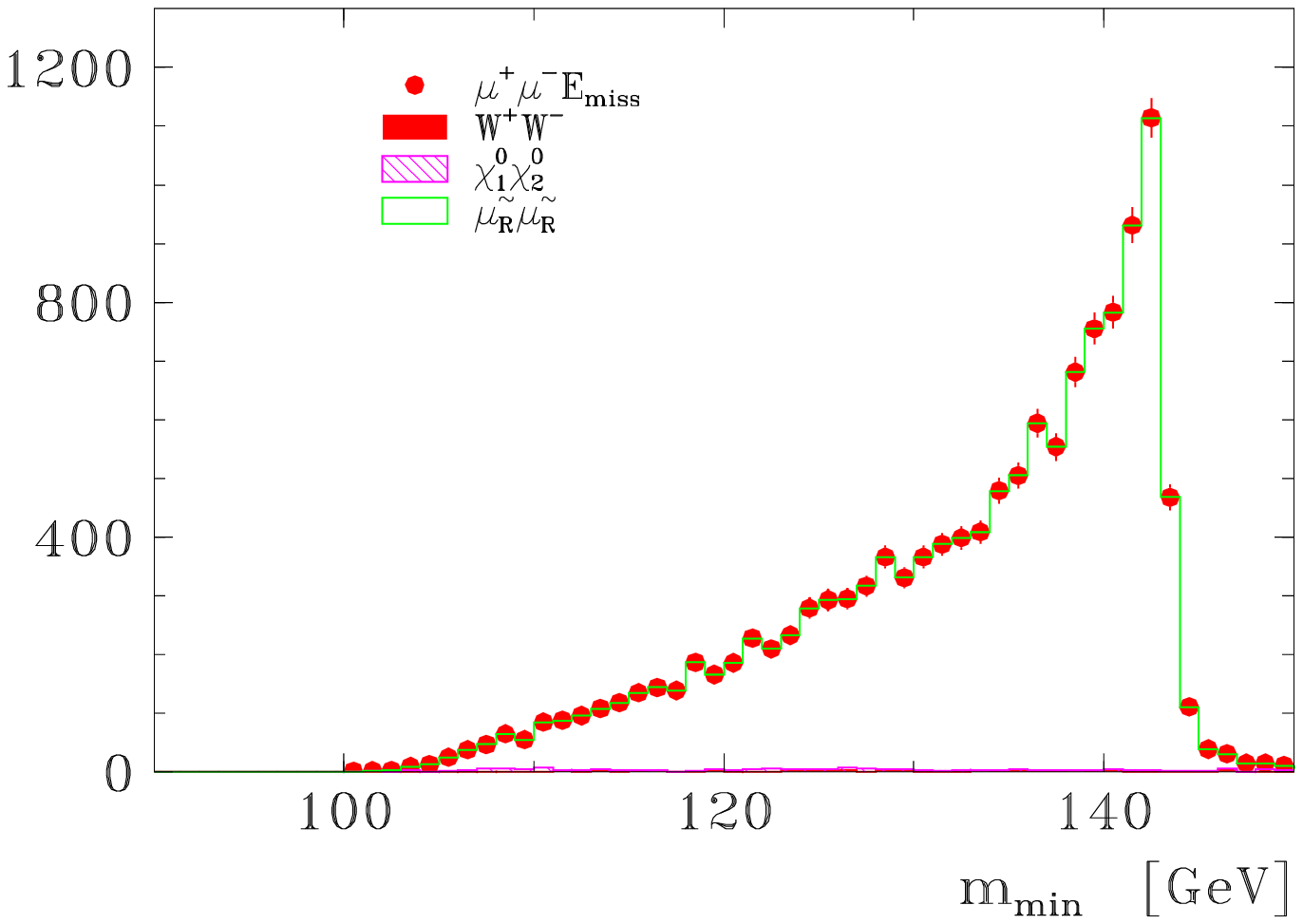,width=.52\textwidth} }
  \caption{Energy spectrum $E_\mu$ of muons (left) and
    minimum mass $m_{\rm min}(\smur)$ (right)
    from the reaction 
    $e^+_L e^-_R\to \smurp  \, \smurm   
    \to \mu^+ \nt_1 \,\, \mu^-\nt_1 $;
    mSUGRA model SPS~1a at $\sqrt{s}=400\;\GeV$ and $\cL=200\;\fbi$}
  \label{smur_mass}
\end{figure}

\paragraph{Mass determination}
Spectra of the muon energy $E_\mu$ and minimum mass 
$m_{\rm min}(\smur)$ are shown in \fig{smur_mass}.
The endpoint energies,
$E_- = 16.570 \pm 0.040\;\GeV$ and
$E_+ = 93.30 \pm 0.22\;\GeV$, 
give strongly correlated $\smur$ and $\nt_1$ masses.
The reconstructed masses from both methods agree well with the input
values. 

\boldmath\bf 
\begin{center}
%\fcolorbox{black}{yellow}{ \color{red} 
\fbox{
\begin{tabular}{l c c}
  method  & $m_{\smur}\ [\GeV]$ & $m_{\nt_1}\ [\GeV]$ \\  \hdick
  $E_\mu$ spectrum      & $ 143.15 \pm 0.17$ 
                        & $ 96.10 \pm 0.21$ \\ 
  $m_{\rm min}(\smur)$ spectrum  & $ 142.98 \pm 0.03$ %~\GeV$ 
                                 & fixed \\[.2em]
  SPS 1a input          & $ 143.0$ & $ 96.0$  
\end{tabular}   }
\end{center}
\unboldmath\rm 

\begin{figure}[htb] %\centering
  \mbox{\hspace{-6mm}
    \epsfig{file=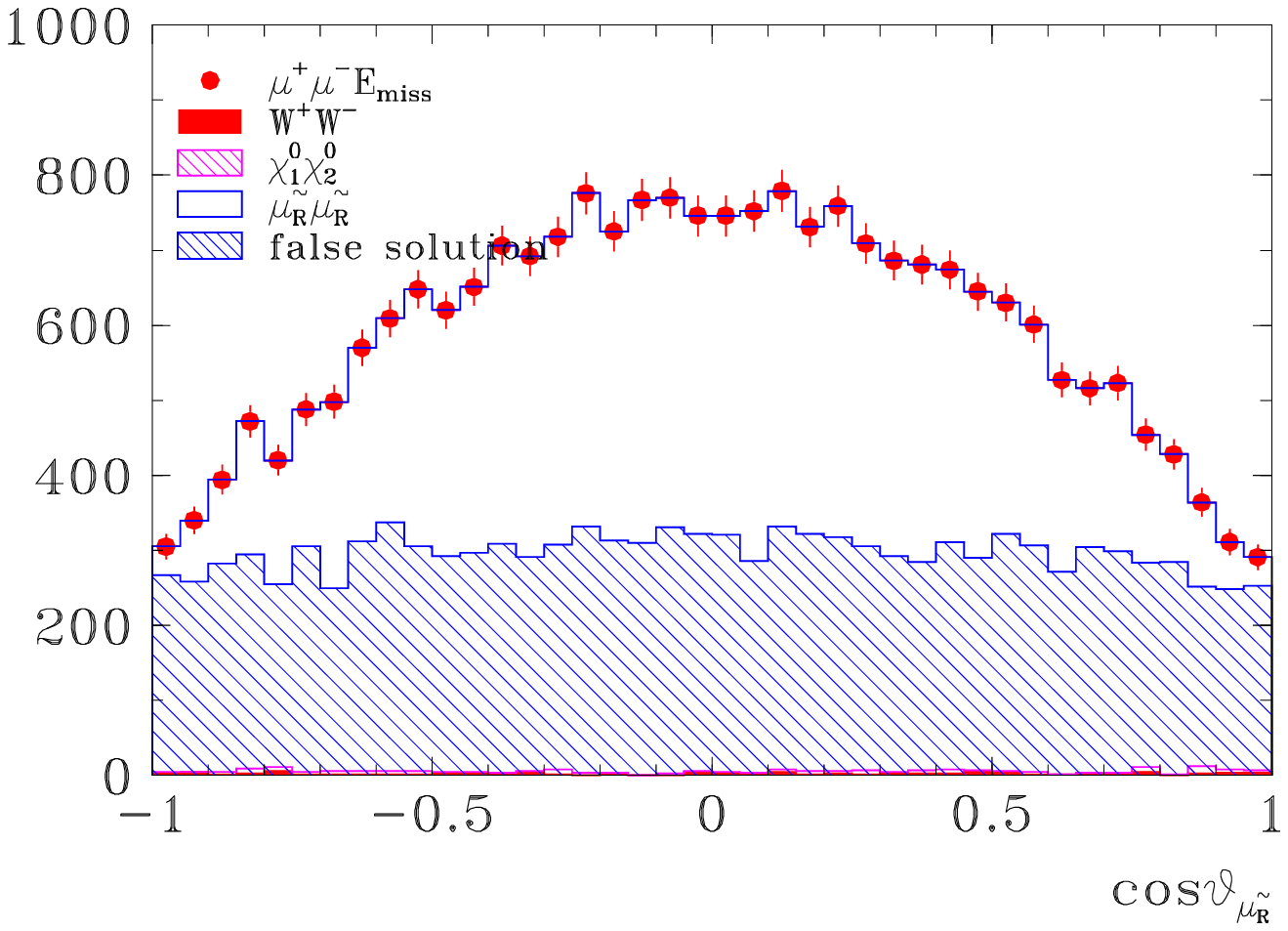,width=.55\textwidth} \hspace{-5mm}
    \epsfig{file=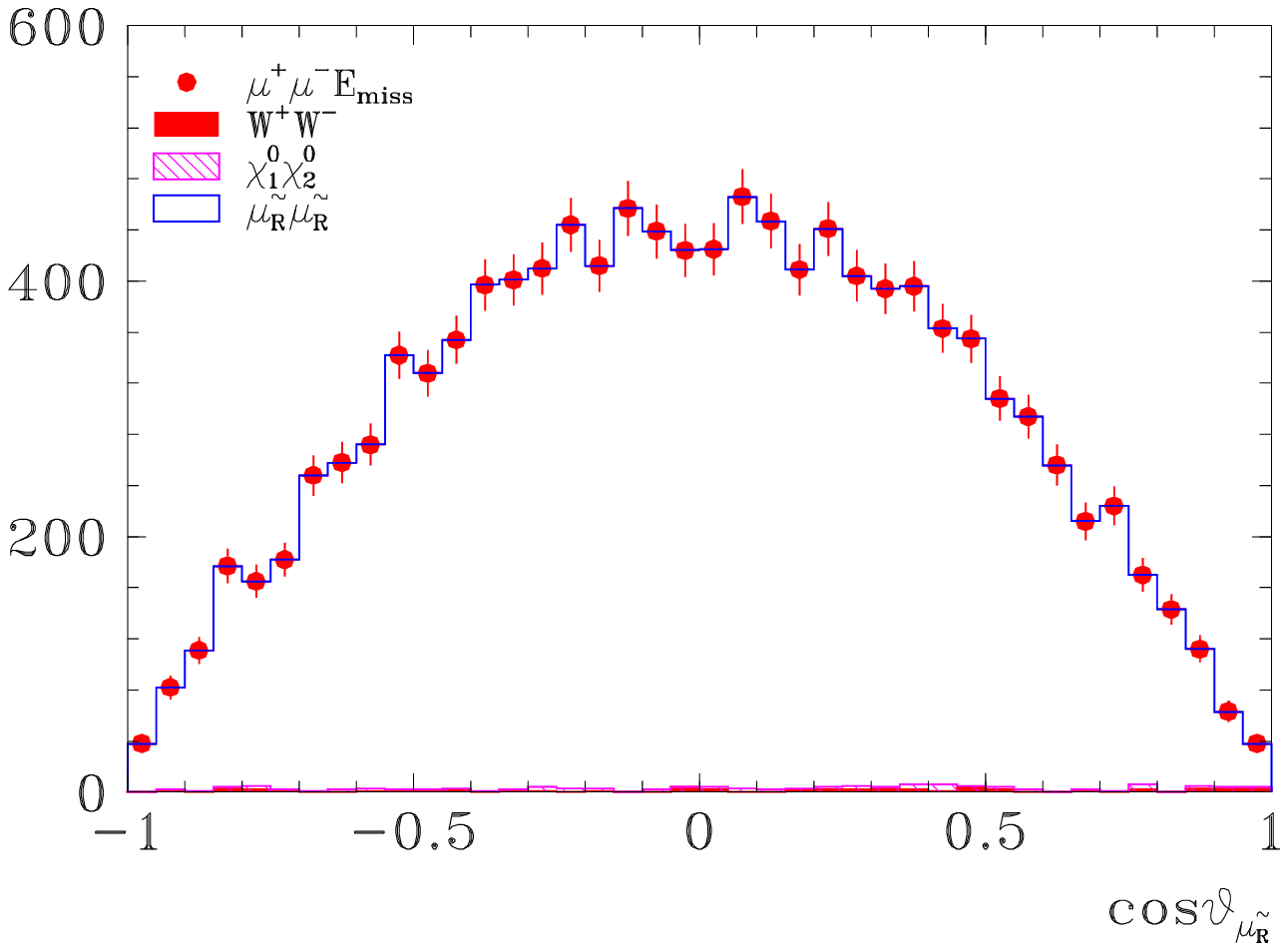,width=.55\textwidth} }
  \caption{Polar angle distribution of $\smur$ with (left) and without (right)
    contribution of false solution in the reaction
    $e^+_L e^-_R\to \smurp  \, \smurm   
    \to \mu^+ \nt_1 \,\, \mu^-\nt_1 $;
    mSUGRA model SPS~1a at $\sqrt{s}=400\;\GeV$ and $\cL=200\;\fbi$}
  \label{smur_cost}
\end{figure}

\paragraph{Spin {\boldmath $J=0$ of $\smur$}}
The polar angle distribution of $\smur\smur$ production is presented
in \fig{smur_cost}.
After background subtraction on finds the typical cross section dependence
\begin{eqnarray*}
  \frac{\rm d \sigma}{\rm d \cos\theta} & \propto &
  \sin^2 \theta
\end{eqnarray*}
as expected for a scalar particle.

\section{Properties of selectrons}

\paragraph{ Selection criteria} 
  $e^+_L e^-_R \to \ser\ser$ at $\sqrt{s}=400~\GeV$
  \footnotesize \scriptsize
  \begin{verbatim} 
 nr  cut                              min       max      eff
 ------------------------------------------------------------
  0. n_ch, n_gamma in accept.           2         0     0.742
     n_charge in acceptance             2         2     0.984
     n_gamma  in acceptance           5.0         0     0.780
     n_veto   in acceptance           5.0         0     0.960
  1. Q*cos_theta_l                 -0.900     0.750     0.586
  2. delta phi_(l+,l-) (deg)                  160.0     0.484
  4. p_T event (GeV)                 10.0               0.484
  5. p_miss cos_theta_(l+l-)                  0.900     0.468
  7. lepton energy (GeV)              2.5     150.0     0.468
 10. m_recoil + 0.4*m_ll (GeV)      240.0               0.468

  efficiency     0.4675    

  lepton energy  E_min 16.52       E_max 93.30    
  \end{verbatim}
  \normalsize

\begin{figure}[htb] \centering
  \mbox{\hspace{-5mm}
  \epsfig{file=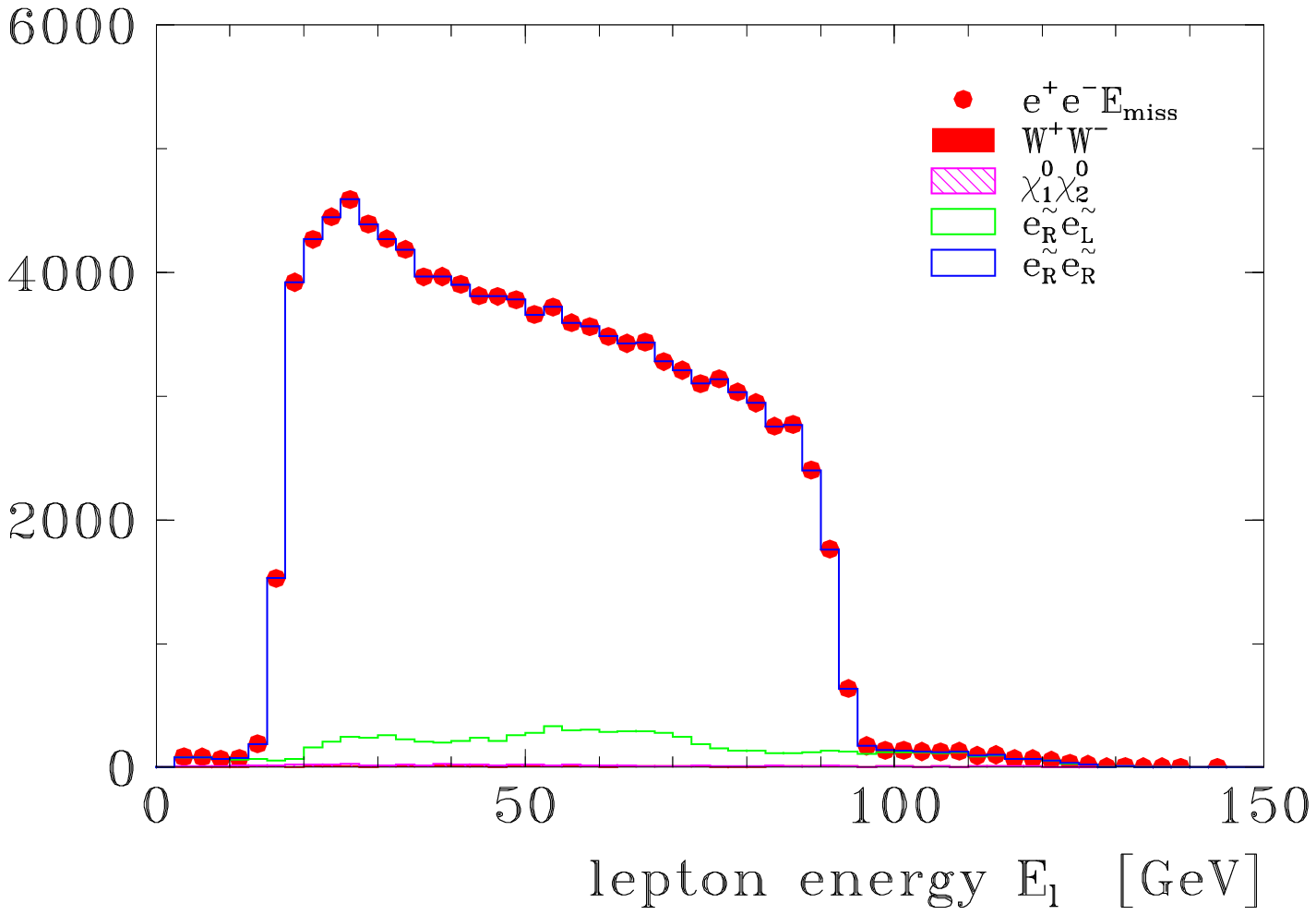,width=.52\textwidth} 
  \epsfig{file=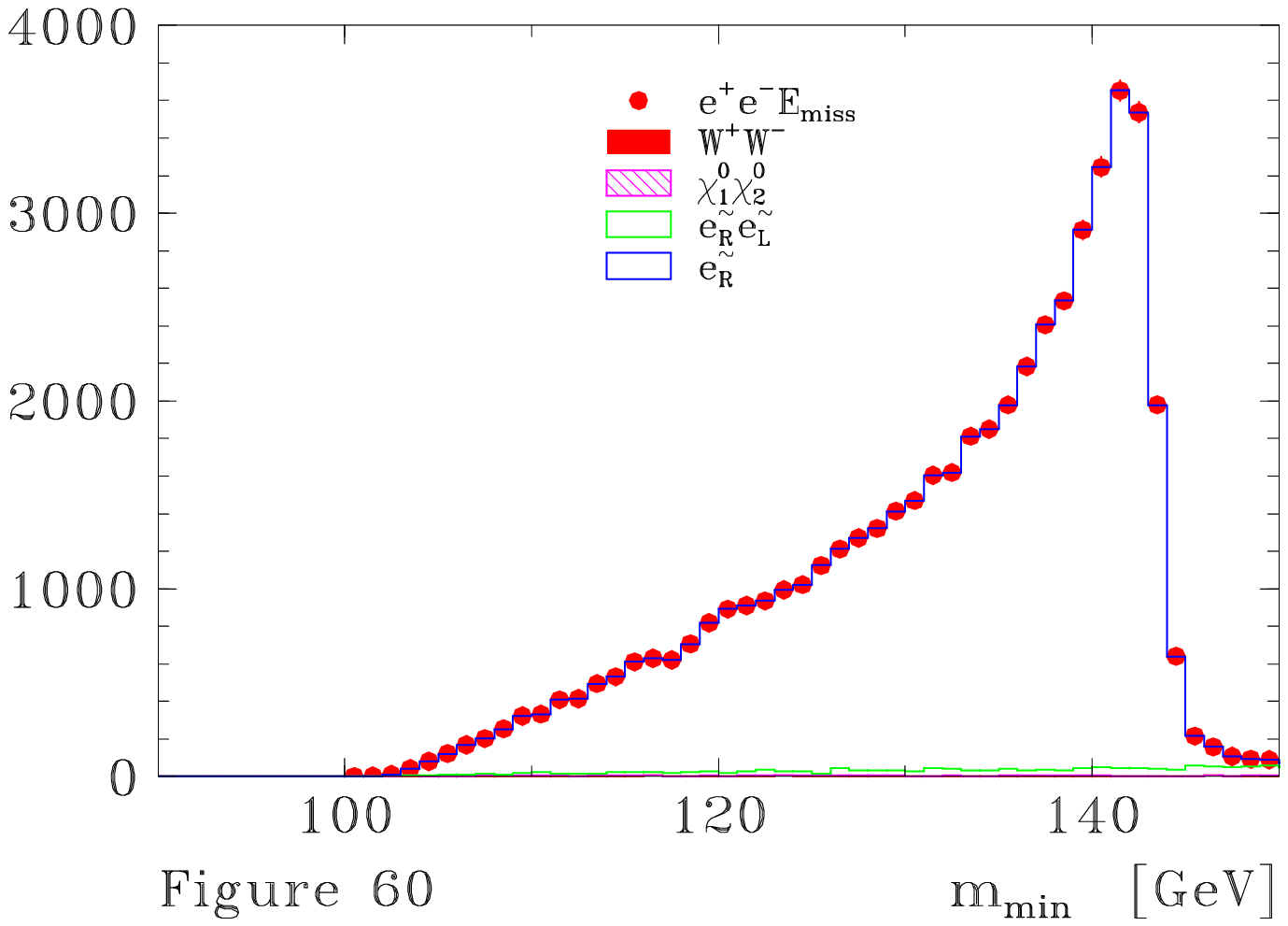,width=.52\textwidth} }
  \caption{Energy spectrum $E_e$ of electrons (left) and
    minimum mass $m_{\rm min}(\ser)$ (right)
    from the reaction 
    $e^+_L e^-_R\to \serp  \, \serm   
    \to e^+ \nt_1 \,\, e^-\nt_1 $;
    mSUGRA model SPS~1a at $\sqrt{s}=400\;\GeV$ and $\cL=200\;\fbi$}
  \label{ser_mass}
\end{figure}

\paragraph{Mass determination}
Spectra of the electron energy $E_e$ and minimum mass 
$m_{\rm min}(\ser)$ are shown in \fig{ser_mass}.
The endpoint energies,
$E_- = 16.528 \pm 0.020\;\GeV$ and
$E_+ = 93.34 \pm 0.11\;\GeV$, 
give strongly correlated $\ser$ and $\nt_1$ masses.
The reconstructed masses from both methods are in agreement with the input
values. 

\boldmath\bf 
\begin{center}
%\fcolorbox{black}{yellow}{ \color{red} 
\fbox{
\begin{tabular}{l c c}
  method  
                      & $m_{\ser}\ [\GeV]$ & $m_{\nt_1}\ [\GeV]$ \\  \hdick
  $E_e$ spectrum      & $ 142.99 \pm 0.08$ 
                      & $ 96.05 \pm 0.10$ \\
  $m_{\rm min}(\ser)$ spectrum  & $ 142.97 \pm 0.02$ %~\GeV$ 
                                 & fixed \\[.2em]
  SPS 1a input          & $ 143.0$ & $ 96.0$  
\end{tabular}   }
\end{center}
\unboldmath\rm 

\begin{figure}[htb] %\centering
  \mbox{\hspace{-6mm}
    \epsfig{file=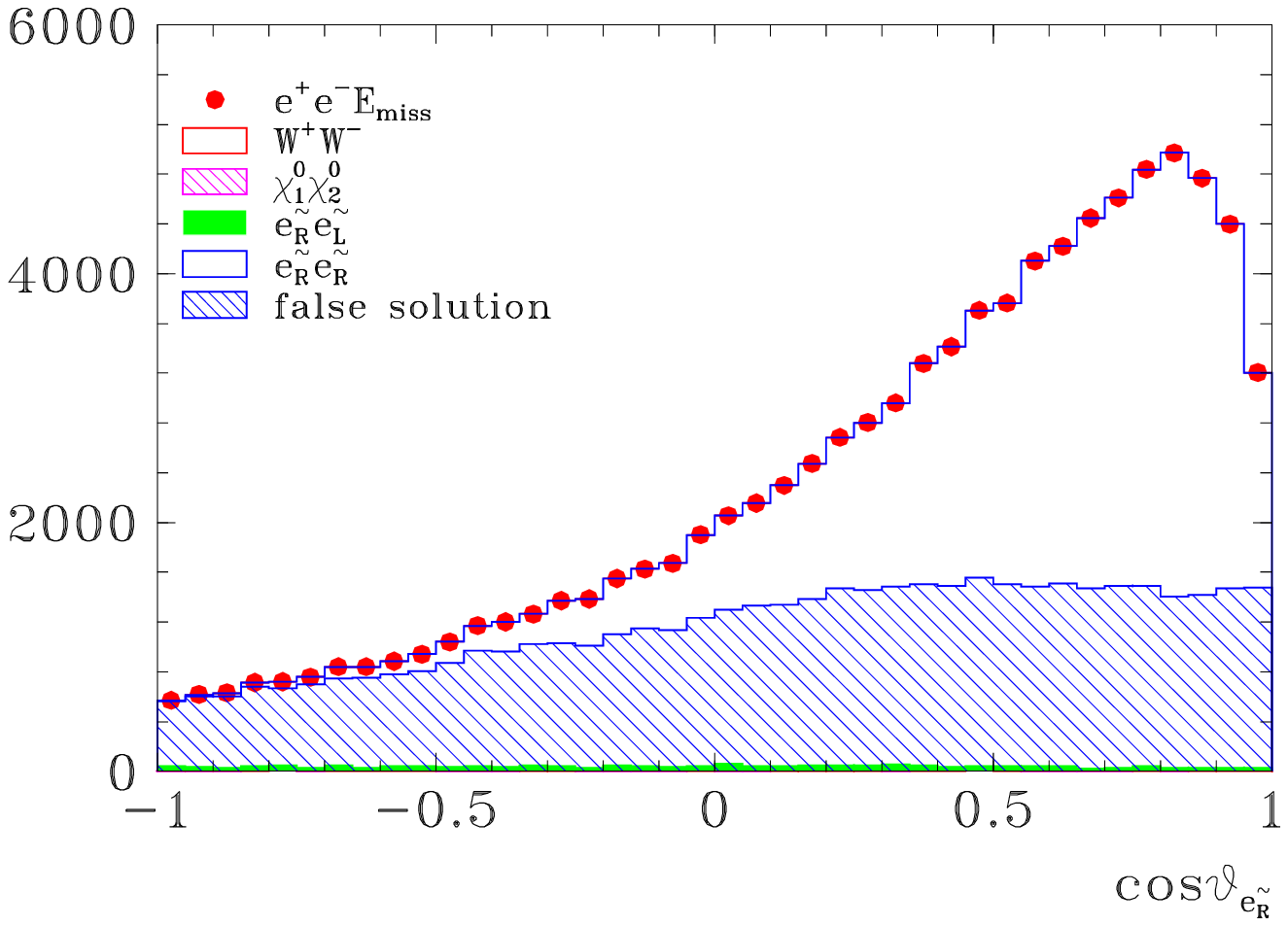,width=.55\textwidth} \hspace{-5mm}
    \epsfig{file=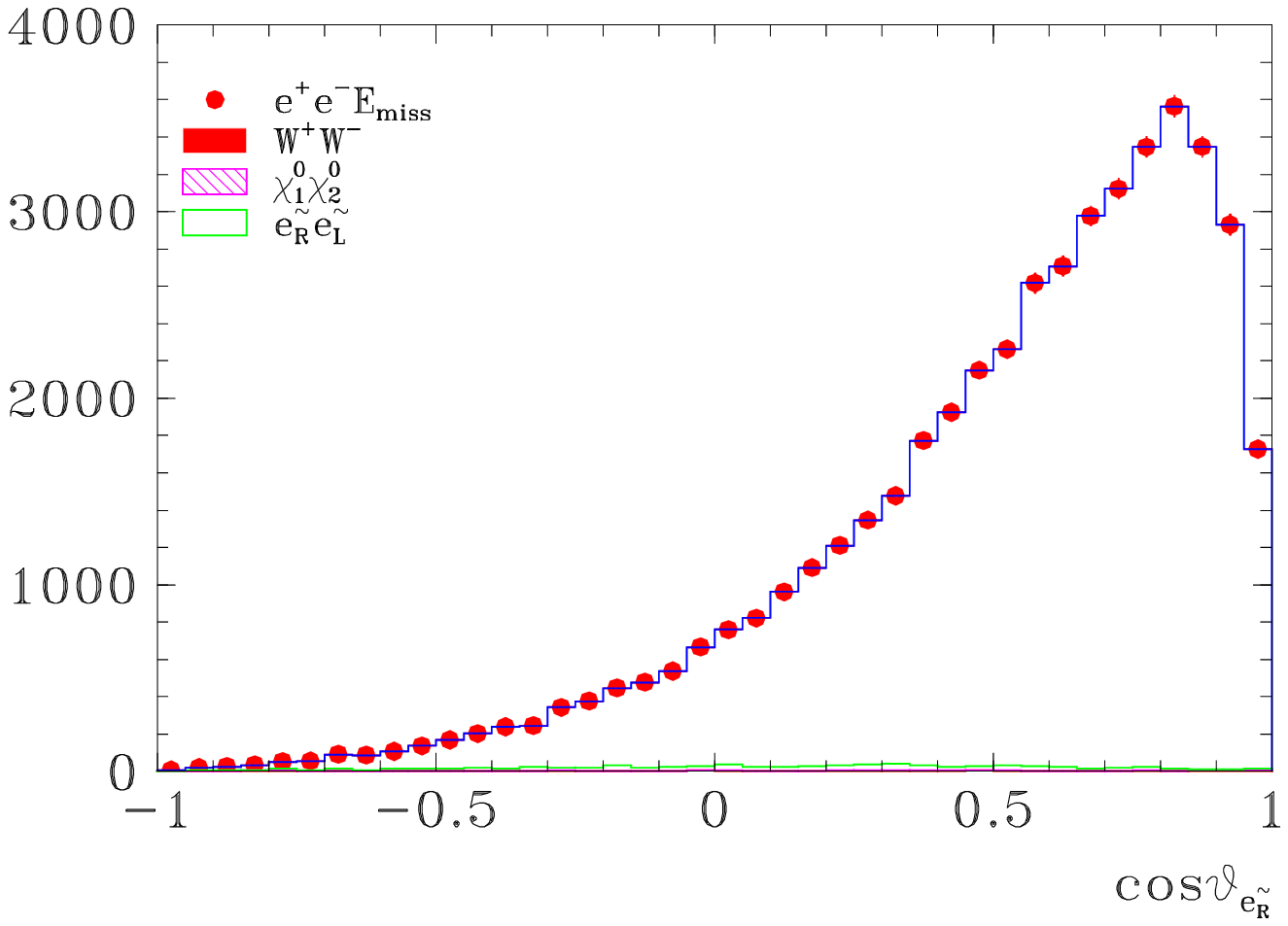,width=.55\textwidth} }
  \caption{Polar angle distribution of $\ser$ with (left) and without (right)
    contribution of false solution
    in the reaction
    $e^+_L e^-_R\to \serp  \, \serm   
    \to e^+ \nt_1 \,\, e^-\nt_1 $;
    mSUGRA model SPS~1a at $\sqrt{s}=400\;\GeV$ and $\cL=200\;\fbi$}
  \label{ser_cost}
\end{figure}

\paragraph{Angular distribution of {\boldmath $\ser$}}
The polar angle distribution of $\ser\ser$ production is presented
in \fig{ser_cost}.
After background subtraction on finds the typical 
behaviour of $t$-channel neutralino exchange, given by the
approximate matrix element 
\begin{eqnarray*}
     {\cal M}  & \sim &  \beta\sin\theta
    \left [ 1 - \frac{4\,Y^2_{\tilde B}}
      {1-2\,\beta\,\cos\theta + \beta^2 + 4\,M_1^2/s} \right ]  \ .
\end{eqnarray*}
The differential cross section %measurement 
can be used to determine the Yukawa
coupling $ Y_{\tilde B}= \hat g'_{\tilde B \se e} / g'_{\gamma e e} $
%\simeq 1.00\pm 0.01$
and the gaugino mass parameter $M_1$ at the percent level.

\begin{figure}[htb] \centering
    \epsfig{file=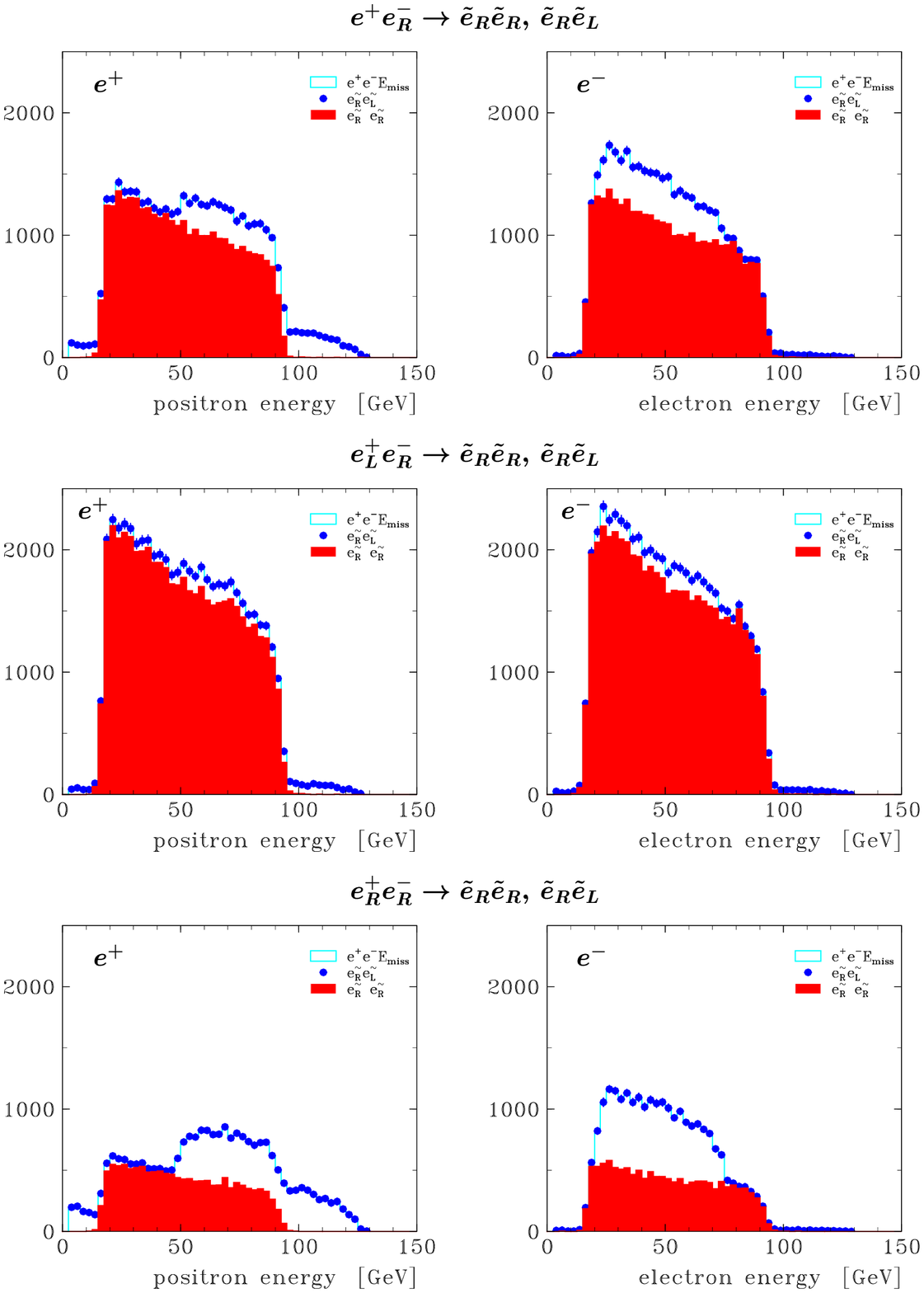,%
      bbllx=50pt,bblly=0pt,bburx=600pt,bbury=740pt,clip=,%
      width=1.\textwidth} 
  \caption{$e^\pm$ energy spectra from
    $\ee\to\ser\ser + \ser\sel$  %and $\ee\to\ser\sel$ 
    production with polarisations
    $\cP_{e^-}=+0.8$, $\cP_{e^+}$ variable 
    for SPS~1a scenario assuming $\sqrt{s}=400\,\GeV$ and
    $\cL=200\,\fbi$}
  \label{prall_serl}
\end{figure}

\begin{figure}[htb] \centering
    \epsfig{file=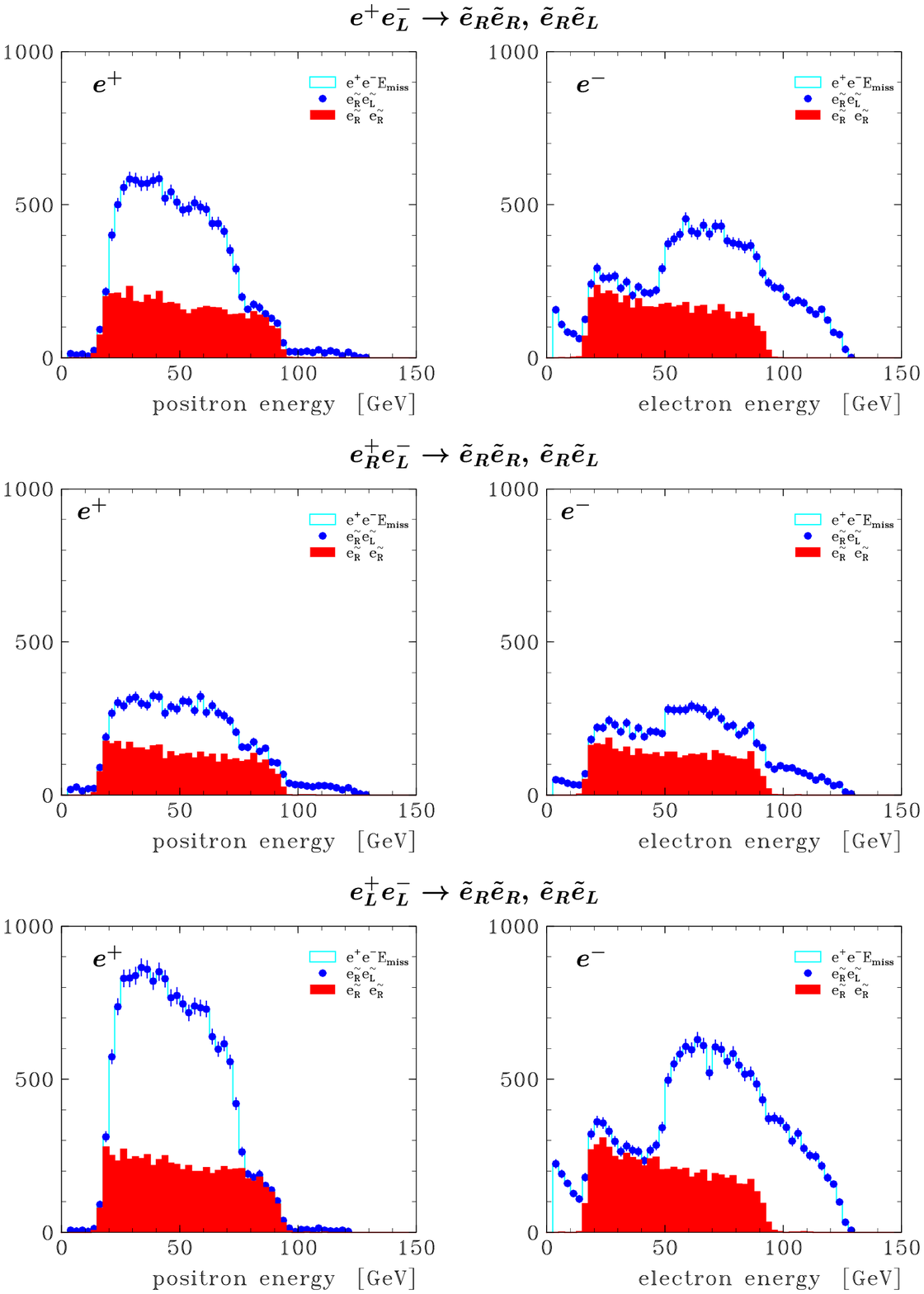,%
      bbllx=50pt,bblly=0pt,bburx=600pt,bbury=740pt,clip=,%
      width=1.\textwidth} 
  \caption{$e^\pm$ energy spectra from
    $\ee\to\ser\ser + \ser\sel$ %and $\ee\to\ser\sel$
    production with polarisations
    $\cP_{e^-}=-0.8$, $\cP_{e^+}$ variable 
    for SPS~1a scenario assuming $\sqrt{s}=400\,\GeV$ and
    $\cL=200\,\fbi$}
  \label{plall_serl}
\end{figure}

\paragraph{Chiral quantum numbers}
A unique determination of the selectron $ L / R $ quantum numbers 
is offered by the reaction $\ee \to \ser\sel$,
which proceed via pure $t$-channel $\nt$ exchange. 
The `chiral flow' at the $e \nt \se$ vertex associates the incoming
electron/positron with its superpartner: 
$e^-_L \to \sel^-$, $e^-_R \to \ser^-$ and
$e^+_L \to \ser^+$, $e^+_R \to \sel^+$.
Measuring the energy spectra of the decay $e^-$ and $e^+$ separately
allows for a clear distinction between $\ser$ and $\sel$.
Moreover, by choosing appropriate beam polarisations,
{\em e.g.} `odd' combinations 
$e^+_R e^-_R \to \sel^+ \ser^-$ or $e^+_L e^-_L \to \ser^+ \sel^-$,
one suppresses the $s$-channel
$\gamma , Z$ amplitudes of $\ser\ser$ and other background
and enhances the $\ser\sel$ production considerably,
as illustrated in \fig{prall_serl} and \fig{plall_serl}.
Contributions from $\ser\ser$ are identical in both $e^\pm$ spectra
and can be readily subtracted since $m_{\ser}$ is known.
A measurement of the $\sel$ mass of 
$m_{\sel} = 202.1 \pm 0.5\;\GeV$ appears feasible.
Similar studies are reported in \cite{dima}.

\section{Properties of staus}

\paragraph{Parameters}
$L - R $ mixing in third generation, access to $\tan\beta$ and
$A_\tau$~\cite{nojiri,boos}
\begin{eqnarray}
         \left(\begin{array}{c}
           \stau_1 \\ \stau_2
               \end{array}\right)                   
             & = &
         \left(\begin{array}{cc}
     \ \cos\tstau & \sin\tstau \\
      -\sin\tstau & \cos\tstau
               \end{array}\right)
         \left(\begin{array}{c}
           \stau_L \\ \stau_R
               \end{array}\right)      \\[1ex]             
  \tan 2\,\theta_{\stau}&=&\frac{-2\,m_\tau\,(A_\tau-\mu\tan\beta)}
           {m^2_{\stau_R} - m^2_{\stau_L}}
\end{eqnarray}
$\tau$ polarisation $\cP_\tau$ in decay $\stau_1 \to \tau \nt_1$
\qquad $\cP_\tau = \cP_\tau(\tstau, \tan\beta, \nt_1\;{\rm mixing})$

\paragraph{Selection criteria}  $e^+_Le^-_R\to\stau^+_1\stau^-_1$ 
                               at $\sqrt{s} = 400\;\GeV$ 
 \footnotesize \scriptsize
 \begin{verbatim}
 nr  cut                              min       max      eff       
 ------------------------------------------------------------
  0. n_charge, n_veto              2    6         1     0.931
  1. n_jets                             2         4     0.834
  2. n_taus                             2         4     0.829
  3. n_leptons                          0         2     0.780
  4. E_jet (GeV)                      2.5     100.0     0.739
  5. Q*cos_theta_jet               -0.950     0.750     0.597
  8. mass tau_jet (GeV)                       2.000     0.547
 11. delta phi_(l,di-jet) (deg)               160.0     0.426
 12. cos theta_P_miss                          0.95     0.424
 13. p_T (GeV)                       10.0     200.0     0.400

  efficiency SUSY      0.3999    

 tau decay statistics
 -------------------- 
             e      mu      pi     rho    3_pi     sum 
 e           0     510     529    1097    1125    3261
 mu          0       0     470    1040    1121    2631
 pi          0       0     200     783     819    1802
 rho         0       0       0     821    1786    2607
 3_pi        0       0       0       0     895     895
 sum         0     510    1199    3741    5746   11196

  lepton energy       E_min 12.19       E_max 83.81    
 \end{verbatim}
 \normalsize 
Leptonic $\tau$ decays not useful and discarded due to large $WW$ background \\
Analysed decay modes \& branching ratios \\ \phantom{xxxxx}
$\cB(\tau\to\pi\nu_\tau) = 0.111, \ \cB(\tau\to\rho\nu_\tau) = 0.254, \
 \cB(\tau\to3\pi\nu_\tau) = 0.194$

\paragraph{{\boldmath $m_{\stau_1}$ from $\tau$ decays} }
$\tau \to \rho\; \nu_\tau \to \pi^\pm\pi^0\; \nu_\tau$ and
$\tau \to 3 \pi\; \nu_\tau \to \pi^\pm\pi^\pm\pi^\mp\; \nu_\tau
                           + \pi^\pm\pi^0\pi^0\; \nu_\tau $ \\
$E_\rho$, $E_{3\pi}$ sensitive to $m_{\stau_1}$,
almost independent of $\cP_\tau$ 

\begin{figure}
\setlength{\unitlength}{1mm}  
\begin{picture}(150,58)     
  \put(-8,-5){ 
    \put(0,0){
      \epsfig{file=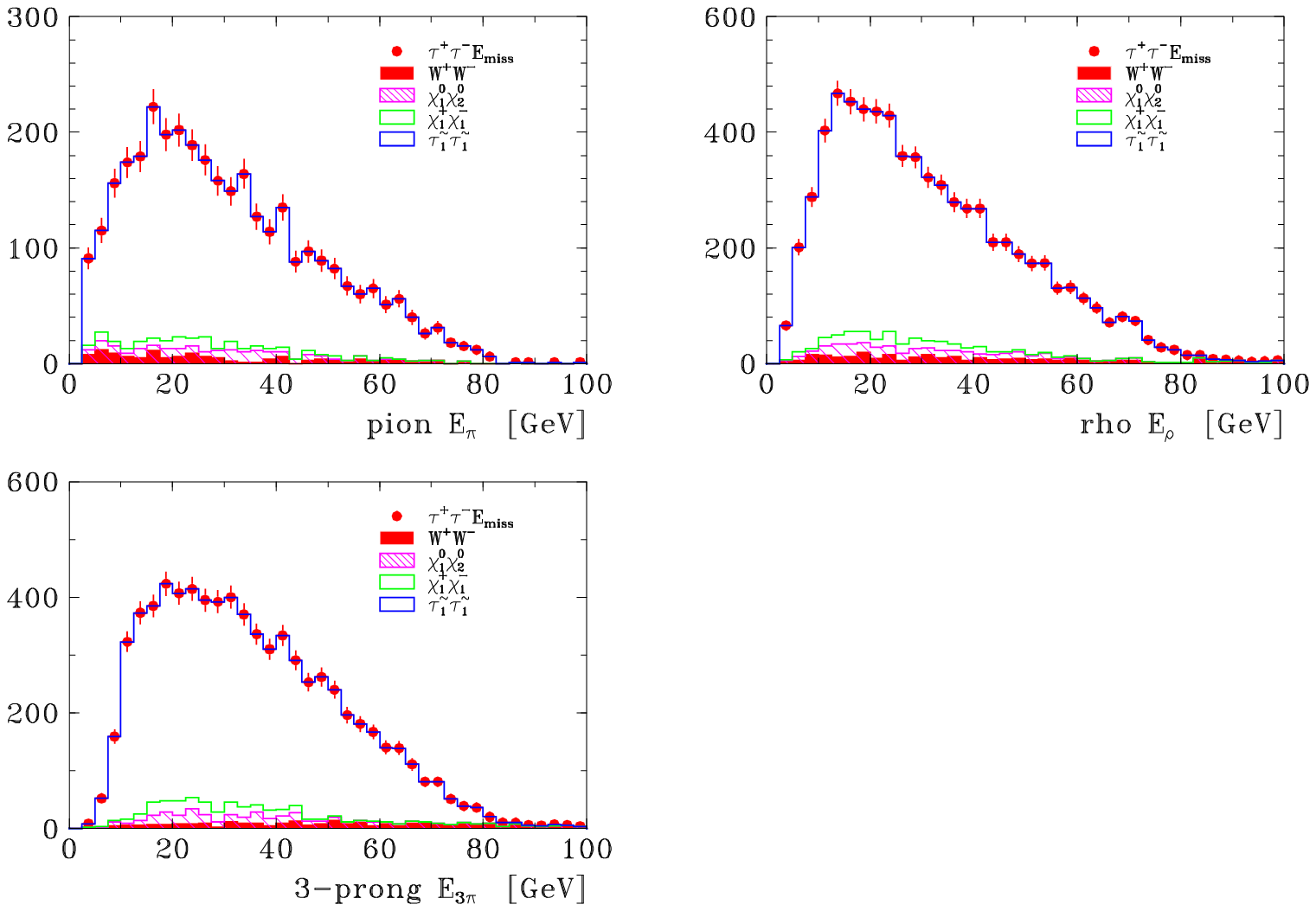,%
        bbllx=315pt,bblly=385pt,bburx=530pt,bbury=545pt,clip=,%
        width=.52\textwidth} 
      \epsfig{file=sps1.stau1.epirho.eps,%
        bbllx=85pt,bblly=230pt,bburx=300pt,bbury=390pt,clip=,%
        width=.52\textwidth} }   
    {\color{blue}
      \put( 16,52){$\tau\to\rho\nu$}  
      \put(100,52){$\tau\to 3\pi\nu$}  }
    }
\end{picture}
  \caption{Hadron energy spectra $E_\rho$ of 
    $\tau\to\rho\nu_\tau$ and $E_{3\pi}$ of
    $\tau\to3\pi\nu_\tau$ decays from the reaction
    $e^+_Le^-_R\to\stau_1^+\stau_1^1\to\tau^+\nt_1\,\,\tau^-\nt_1$
    in SPS~1a scenario assuming $\sqrt{s}=400\,\GeV$ and
    $\cL=200\,\fbi$}
  \label{stau1_mass}
\end{figure}

\noindent
Analysis of $E_\rho$ and $E_{3\pi}$  spectra, shown in \fig{stau1_mass},
assuming $\mnt{1}$ known

\begin{center}
\boldmath\bf
\fbox{
  \begin{tabular}{l c}
    mode      & $\mstau{1}~[\GeV]$ \\ \hdick
    $E_\rho$   & $ 133.3 \pm 0.75$ \\
    $E_{3\pi}$ & $ 133.2 \pm 0.30$ \\ \hline
    mean       & $ 133.2 \pm 0.30$ \\ 
    SPS 1a input & $ 133.2$
  \end{tabular} }   
\unboldmath\rm
\end{center}

\paragraph{{\boldmath $\tstau$} from cross section }
A cross section measurement of
$\sigma_{\stau_1\stau_1} = 137.8 \pm 2.2_{\rm stat} 
                                 \pm 1.4_{\rm sys}~\fb$
(incl. ISR \& BS) provides a $\stau$ mixing angle of
$\cos 2\,\tstau =-0.84 \pm 0.04$, see \fig{stau1_cos2t}

\paragraph{{\boldmath $\tau$ polarisation $\cP_\tau$}}
$E_\pi$ from $\tau\to\pi\,\nu_\tau$  
and $z_\pi=E_\pi/E_\rho$ from  $\tau\to\rho\nu_\tau\to
\pi\pi^0\nu_\tau$ \\
$E_\pi$, $z_{\pi}$ sensitive to  $\cP_\tau$,
almost independent of $m_{\stau_1}$ \\[1ex]
$\tau$ is self-analysing system -- spin correlations   \\[.3em]
$\tau^-_R \to \pi^- \nu$ \quad $\cP_\tau = +1$ 
\qquad  $\pi^-$ emitted  in forward direction 
\qquad $E_\pi \nearrow$ \\
$\tau^-_L \to \pi^- \nu$ \quad  $\cP_\tau = -1$
\qquad  $\pi^-$ emitted  in backward direction 
\quad\, $E_\pi \searrow$
\\[1ex]
$\tau^-_R \to \rho_L^- \nu$ \quad $\cP_\tau = +1$ 
\qquad  ${\rm d}\sigma/{\rm d} z_\pi \propto (2 z_\pi-1)^2$
\quad \ asymmetric sharing $z_\pi \to 0 \, / \,  1$ \\
$\tau^-_L \to \rho_T^- \nu$ \quad  $\cP_\tau = -1$
\qquad ${\rm d}\sigma/{\rm d} z_\pi \propto 2 z_\pi(1-z_\pi)$
\quad equal sharing $z_\pi \to 0.5$

\begin{figure}
\setlength{\unitlength}{1mm}  
\begin{picture}(150,58)     
  \put(-8,-5){ 
    \put(0,0){
      \epsfig{file=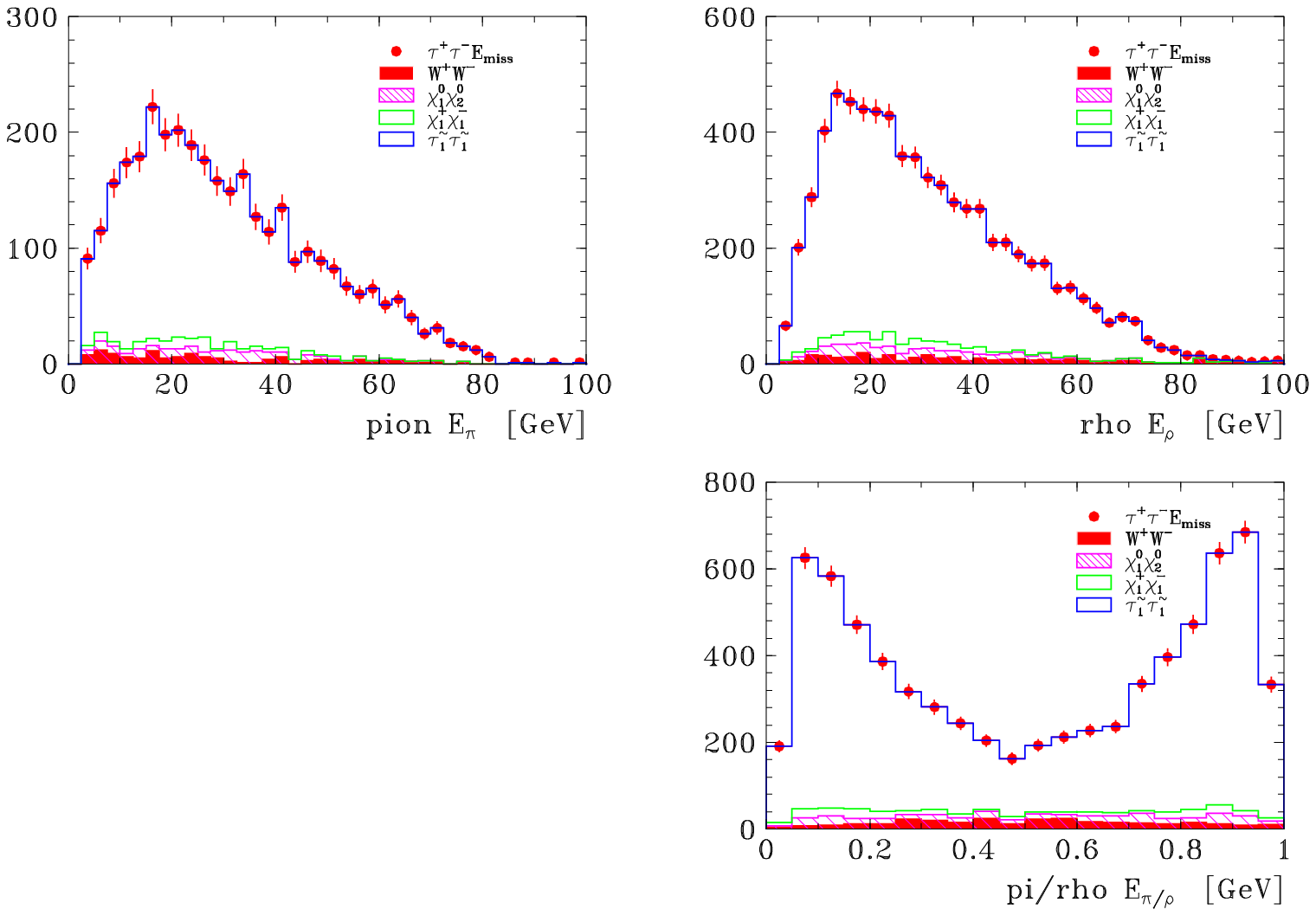,%
        bbllx=85pt,bblly=385pt,bburx=300pt,bbury=545pt,clip=,%
        width=.52\textwidth} 
      \epsfig{file=sps1.stau1.epol.eps,%
        bbllx=315pt,bblly=230pt,bburx=530pt,bbury=390pt,clip=,%
        width=.52\textwidth} }   
    {\color{blue}
      \put(16,52){$\tau\to\pi\nu$}  
      \put(100,52){$\rho^\pm\to\pi^\pm\pi^0$}  }
    }
\end{picture}
  \caption{Pion energy spectrum $E_\pi$ of $\tau\to\pi\nu_\tau$
    and ratio $z_\pi=E_\pi/E_\rho$ of
    $\tau\to\rho\nu_\tau$ decays from the reaction
    $e^+_Le^-_R\to\stau_1^+\stau_1^1\to\tau^+\nt_1\,\,\tau^-\nt_1$
    in SPS~1a scenario assuming $\sqrt{s}=400\,\GeV$ and
    $\cL=200\,\fbi$}
  \label{stau1_pol}
\end{figure}

\noindent
Analysis of $E_\pi$ and $z_\pi$ spectra, shown in \fig{stau1_pol},
$\mnt{1}$ assumed to be known
\begin{center} \boldmath\bf
  \fbox{
    \begin{tabular}{l c}
      mode      
                     & $\cP_\tau$ \\ \hdick
      $E_\pi$        & $ 0.98 \pm 0.045$ \\ 
      $z_\pi=E_\pi/E_\rho$ & $ 1.02 \pm 0.038$ \\  \hline        
      mean      & $ 1.00 \pm 0.035$ \\
      SPS 1a input      & $ 0.981$ \\
      \end{tabular} }  
  \unboldmath\rm
\end{center}

\noindent
A polarisation measurement of $\cP_\tau = 1.00 \pm 0.035$ 
is not sensitive enough to determine $\tan\beta$ in SPS~1a, where one has
$\delta \cP_\tau/\delta\tan\beta = 0.3\% $

\section{Summary}

Simulations of slepton production 
$\ee\to\ser\ser,\; \smur\smur,\; \stau_1\stau_1$ 
in the SPS~1a mSUGRA scenario under realistic experimental conditions assuming
$\sqrt{s} = 400~\GeV$, $\cL = 200~\fbi$ and beam polarisations
$\cP_{e^-} = +0.8$, $\cP_{e^+} = -0.6$ are presented. 
The slepton and neutralino $\nt_1$ masses 
as well as the polarisation $\cP_{\stau_1\to \tau\nt_1}$ and 
mixing parameter $\tstau$ of the third
generation can be accurately determined with moderate integrated
luminosity. 
Improvements on $m_{\ser}$ and  $m_{\smur}$
are possible if $\delta\mnt{1}$ can be reduced, 
{\em  e.g.} by studying neutralino production and decays.
An alternative method to measure selectron and smuon masses via
cross section threshold scans is discussed in \cite{freitas}.
Other estimates on SPS~1a sparticle masses 
can be found in \cite{dima} and \cite{grannis} 
(not based on detailed simulations).
\begin{center} \boldmath\bf
  \fbox{
    \begin{tabular}{l c c}
      $\ser$    & $m_{\ser} =  142.99 \pm 0.08~\GeV$ 
                & $m_{\nt_1} = 96.05 \pm 0.10~\GeV$  \\
      $\smur$   & $m_{\smur} = 143.15 \pm 0.17~\GeV$ 
                & $m_{\nt_1} = 96.10 \pm 0.21~\GeV$ \\
      $\stau_1$ & $m_{\stau_1} = 133.2 \pm 0.30~\GeV$ & ---  \\
                & $\cos2\,\tstau = -0.84 \pm 0.04$ 
                & $\cP_\tau = 1.00 \pm 0.035 $ 
    \end{tabular} }  
  \unboldmath\rm
\end{center}

\begin{figure}[htb] \centering
  \mbox{\hspace{-6mm}
    \epsfig{file=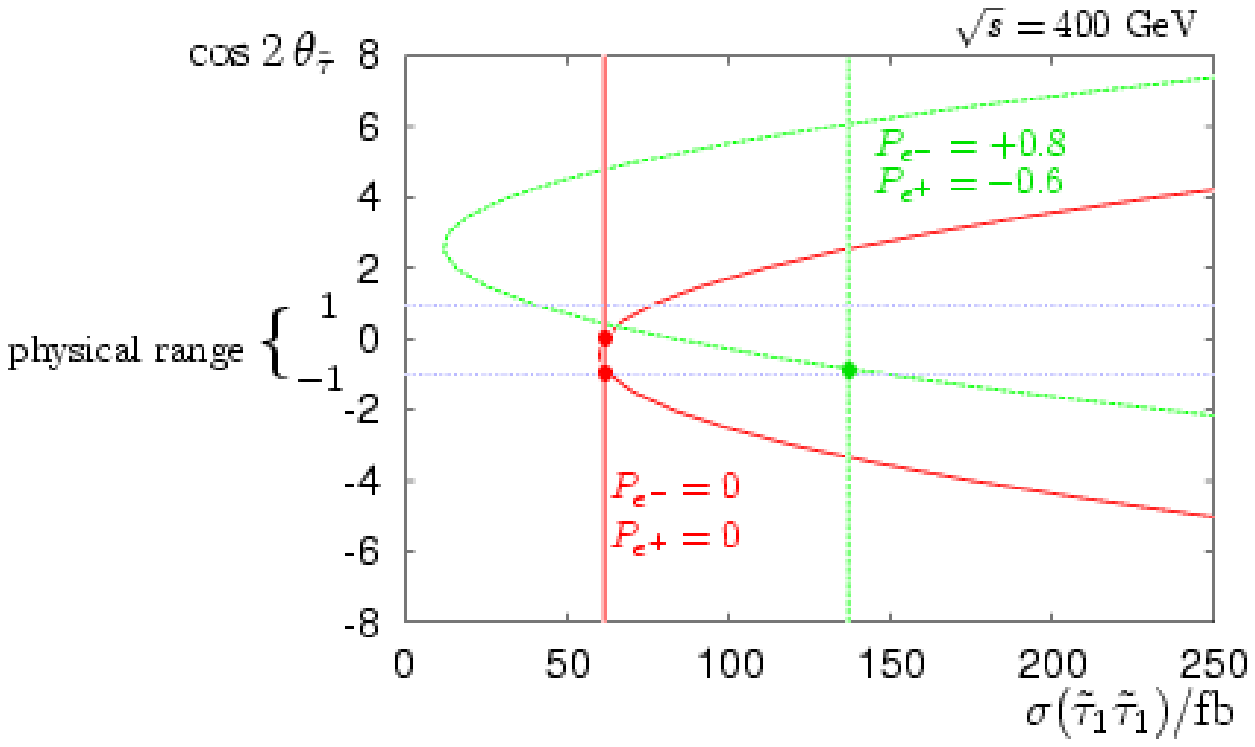,width=.6\textwidth} 
    %\hspace{-5mm}
    %\epsfig{file=sps1_ser_cost,width=.55\textwidth} 
  }
  \caption{Mixing angle angle $\cos 2\tstau$ versus cross section
    $\sigma_{\stau_1\stau_1}$ at $\sqrt{s}=400\;\GeV$ %(left) and 
    %$\tan\beta$ versus $\cP_{\stau_1\to\tau\nt_1}$ (right)
    for mSUGRA scenario SPS~1a}
  \label{stau1_cos2t}
\end{figure}

\paragraph{Acknowledgement}
I want to thank G.~Moortgat-Pick for the stimulating discussions on the
stau sector and for providing the plot of \fig{stau1_cos2t}.

%
%   References 
%

%
%

\clearpage
\begin{figure} \centering
\setlength{\unitlength}{1mm}
\begin{picture}(145,180)(0,0) \boldmath\bf
  \put(0,0){
    \put(100,155){{\Large SPS 1a}}
    \put(20,155){{\large mSUGRA}}
    \put(20,148){{ $m_0 =  100,\, m_{1/2} = 250,\, A_0 = -100 $}}
    \put(20,143){{ $\tan\beta = 10$, sign$\,\mu\, + $}}
    \put(0,10){\epsfig{file=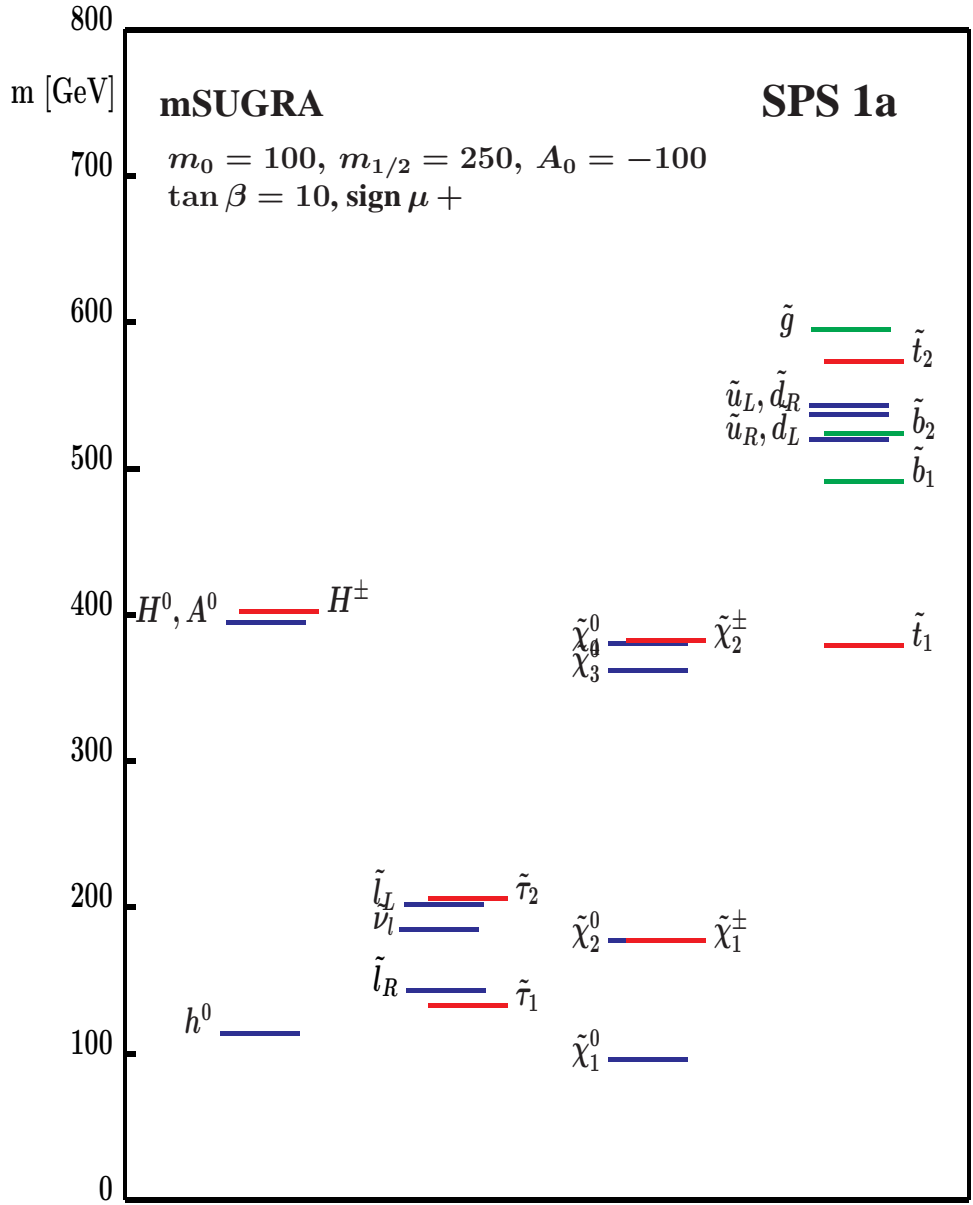,width=0.8\textwidth,
        height=01.\textwidth}}
    }
\end{picture}
\caption{Particle spectrum of mSUGRA benchmark scenario SPS~1a}
\label{mass_spectrum}
\end{figure} 

\clearpage

\clearpage

\clearpage

%\clearpage

\vfill


\begin{thebibliography}{99}
        
\bibitem{tdr} {\sc Tesla} Technical Design Report, DESY 2001-011,
        Part III: {\em Physics at an $e^+e^-$ Linear Collider}
        [hep-ph/0106315],
        Part IV: {\em A Detector for TESLA}.
        %[hep-ph/0106315]
\bibitem{spsmodels} B.C.~Allanach et al., %{\it Snowmass points},  
        Eur. Phys. J. C 25 (2002) 113; \\ % [hep-ph/0202233]; \\
        N.~Ghodbane, H.-U.~Martyn, hep-ph/0201233.
        %B.C.~Allanach et al., Eur. Phys. J. C 25 (2002) 113.
         
\bibitem{pythia} {\sc Pythia}, T.~Sj\"ostrand et al.,
        Comput. Phys. Commun. 135 (2001) 238.
         
\bibitem{circe} {\sc Circe}, T.~Ohl,
        Comput. Phys. Commun. 101 (1997) 269.
         
\bibitem{tauola} {\sc Tauola}, S.~Jadach et al.,
        Comput. Phys. Commun. 76 (1993) 361.
         
\bibitem{simdet} {\sc Simdet}, M.~Pohl, J.~Schreiber,
        DESY-02-061, hep-ex/0206009.
        
\bibitem{feng} J.~Feng, D.~Finell,
        Phys. Rev. D 49 (1994) 2369. %[hep-ph/9310211].
        
\bibitem{dima} M.~Dima et al.,
        Phys. Rev. D 65 (2002) 71701. %[hep-ph/0112017].

\bibitem{nojiri} M.M.~Nojiri,
        Phys. Rev. D 51 (1995) 6281; \\ % [hep-ph/9412374]; % \\
        M.M.~Nojiri, K.~Fujii, T.~Tsukamoto,
        Phys. Rev. D 54 (1996) 6756. % [hep-ph/9606370].

\bibitem{boos} E.~Boos et al., Eur. Phys. J. C 30 (2003) 395.
        %[hep-ph/0303110] %{\it Polarisation in sfermion decays}

\bibitem{freitas} A.~Freitas, A.~von~Manteuffel, 
        proceedings SUSY~02, DESY Hamburg, Germany, 2002,
        hep-ph/0211105.

\bibitem{grannis} P.~Grannis, talk at LCWS~02, Jeju Island, Korea, 2002,
        hep-ex/0211002.

\end{thebibliography}
\end{document}